\documentclass[prl,twocolumn,amsmath,amssymb,superscriptaddress,showpacs]{revtex4-1}
\usepackage{graphicx}
\usepackage{color}

\newcommand{\beq}{\begin{equation}}
\newcommand{\eeq}{\end{equation}}
\newcommand{\be}{\begin{equation}}
\newcommand{\ee}{\end{equation}}
\newcommand{\beqn}{\begin{eqnarray}}
\newcommand{\eeqn}{\end{eqnarray}}
\newcommand{\bea}{\begin{eqnarray}}
\newcommand{\eea}{\end{eqnarray}}
\newcommand{\bearr}{\begin{array}}
\newcommand{\enarr}{\end{array}}

\newcommand{\comment}[1]{}

\begin{document} 

\title{Crisis in time-dependent dynamical systems}
      
\author{Simona Olmi}
\email{ simona.olmi@cnr.it}
\affiliation{Istituto dei Sistemi Complessi, Consiglio Nazionale
delle Ricerche, via Madonna del Piano 10, I-50019 Sesto Fiorentino, Italy}
\affiliation{INFN, Sezione di Firenze, Via Sansone 1, I-50019 Sesto Fiorentino, Italy}

\author{Antonio Politi}
\email{a.politi@abdn.ac.uk}
\affiliation{Institute for Complex Systems and Mathematical Biology  
           University of Aberdeen, Aberdeen AB24 3UE, United Kingdom}
\affiliation{Istituto dei Sistemi Complessi, Consiglio Nazionale
delle Ricerche, via Madonna del Piano 10, I-50019 Sesto Fiorentino, Italy}

\begin{abstract}
Many dynamical systems operate in a fluctuating environment. However, even in low-dimensional setups, transitions
and bifurcations have not yet been fully understood.
In this Letter we focus on crises, a sudden flooding of the phase space due to the crossing of the boundary
of the basin of attraction.  We find that crises occur also in non-autonomous systems
although the underlying mechanism is more complex. We show that in the vicinity of the transition,
the escape probability scales as $\exp[-\alpha (\ln \delta)^2]$, where $\delta$ is the distance from the
critical point, while $\alpha$ is a model-dependent parameter. This prediction is tested and verified in
a few different systems, including the Kuramoto model with inertia, where
the crisis controls the loss of stability of a chimera state. 
\end{abstract}

\maketitle

The study of forced dynamical systems has recently attracted much interest, since
this setup allows analysing complex systems under the assumption that a given 
fraction of the degrees of freedom can be treated as an external noise-like drive.
This is particularly useful in the context of global models of climate evolution, where the
concept of ``pull-back" attractors has been introduced precisely for this goal~\cite{Crauel2015,Ghil2018}.
Another research area where this simplification proves useful is synchronization of ``slave" systems
forced by a ``master", especially in the presence of a dynamic control~\cite{Ramirez2018}.
This includes the study of globally coupled oscillators, where 
the self-determined mean field can, in many respects, be treated as an external drive.

Within the more mathematically oriented community, these dynamical systems are identified as ``non-autonomous".
They undergo similar qualitative changes 
to those exhibited by (time) translationally invariant dynamical systems, when some control
parameters are varied.
However, they are complicate to analyse since the instantaneous configuration is not unique; it depends 
on the (typically unknown) configuration of the master (see~\cite{Clemson2014} for a review).

In this Letter, we focus on a transition extensively investigated in chaotic low-dimensional systems: the ``crisis",
where the attractor reaches (and crosses) the boundary of its basin of attraction, suddenly widening the 
region of phase space explored by the stationary state~\cite{Grebogi1982}.
A preliminary study of a crisis in forced systems has been performed in a model 
of El Ni\~no-Southern Oscillation~\cite{Checkroun2018}.
Here, we investigate this phenomenon in a context where a parameter fluctuates, being the result of an
external nonlinear dynamical process.
Crises in noisy dynamical systems had been extensively studied in the '90's of the previous century, but the analysis
was always focused on theoretical estimates of the escape time in the vicinity of
the deterministic critical point. The goal was achieved by expressing the dependence of the escape time
on a single parameter: the ratio between the distance from the critical point and the noise
amplitude (the latter one being either unbounded as 
in~\cite{Sommerer1991,Sommerer1991b,Franaszek1991,ReimannI}, or bounded as in
\cite{ReimannII,Wackerbauer1999}).
Here, we show that under the assumption of bounded (not necessarily small) fluctuations,
``noise" shifts the position of the transition point and induces an antirely new scaling behavior.

We first investigate the occurrence of a crisis with reference to the second-order Kuramoto model,
commonly used for describing networks of oscillators capable of adjusting their natural frequencies. 
The inclusion of inertia induces a complex collective dynamics, where a fluctuating chimera state 
destabilizes via one such type of crisis. 
Numerical simulations reveal a strong divergence of the escape time, exhibited also
by simplified models such as the modulated H\'enon- and logistic-map.
Thanks to a further simplification, we are able to derive analytical formulas, which are found
to reproduce the behavior of various dynamical models.

\begin{figure}
\includegraphics[width=0.5\textwidth,clip]{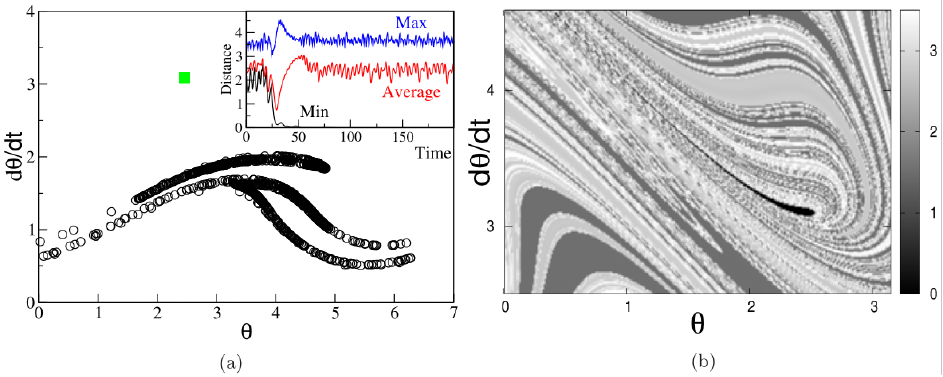}
\caption{(a) Snapshot of the Kuramoto model in the phase plane $(\dot{\vartheta},\vartheta)$. 
The green square denotes the cluster position, while the open circles denote the dust.
($f_{cl}=0.393$). Inset: time dependence of the dust-cluster distance in the presence of a migration event. 
The blue (red) curve identifies the maximal (average) Euclidean distance between the dust and the cluster.  
The black curve identifies the distance between the cluster and the nearest dust oscillator. 
Around t=100 the minimal distance virtually vanishes, indicating the occurrence of a migration event. 
(Initially, $f_{cl}=0.465.$).
(b)  Basin of attraction of the cluster. 
The greyscale in ${\bf P}=(\vartheta, \dot{\vartheta})$ identifies the Euclidean distance 
of a probe oscillator (initially in ${\bf P})$ from the cluster after a time $t_e=100$.
Initial conditions are varied in a grid of size 0.01 in both directions.
The cluster is initially in $(2.48160,3.12)$ and $f_{cl}= 0.5$.
}
\label{fig:kuramoto}
\end{figure}

\paragraph{Globally coupled Kuramoto rotors.}
Here we consider a network of $N$ identical, symmetrically coupled rotators, each characterized by a phase $\vartheta_i$ 
and a frequency $\dot{\vartheta}_i$.
The phase $\vartheta_i$ of the $i$th oscillator evolves according to the differential equation
\begin{equation}\label{dynamics-eq}
 m\ddot{\vartheta_i}(t) + \dot{\vartheta_i}(t) = \frac{K}{N}\sum_{j}^N \sin\left(\vartheta_j(t)-\vartheta_i(t) +\gamma\right) \,,
\end{equation}
where $m=6$ is the inertia ($1/m$ plays the role of dissipation) and $\gamma=1.6$ is a fixed phase lag.
The oscillators are homogeneously coupled (we set $K=6$, without loss of generality).
Symmetry may spontaneously break, splitting
the entire population in groups which exhibit a different behavior.
Particularly interesting is the case when a {\it dust} of non synchronized units coexists with 
 one or more {\it clusters} of perfectly synchronized oscillators.
This regime, called chimera state, has been explored in several 
setups~\cite{kuramoto2002coexistence, abrams2004chimera, omel2008chimera, pikovsky2008partially, laing2009chimera, olmi2011collective, scholl2016synchronization, zakharova2020chimera}; it emerges also in the Kuramoto model with 
inertia~\cite{jaros2015chimera, olmi2015intermittent, olmi2015chimera}.
An exemplary  chimera snapshot is presented in Fig.~\ref{fig:kuramoto}(a), where the empty black dots identify the dust
(composed of 1214 oscillators), while the green square represents a cluster (composed of 786 oscillators).
Since all oscillators are identical, the stability of such regimes can be fruitfully investigated by focusing on the
response of a single oscillator to a given mean field. This way, the problem is recognized as an instance
of stability assessment of a time-dependent (non-autonomous) dynamical system.
This is transparent once  we rewrite the evolution equations in terms of the Kuramoto order
parameter 
\begin{equation}\label{EQ:order_parameter}
R(t) \mathrm{e}^{i\Phi(t)} = \frac{1}{N_{d}} \sum\limits_{j=1}^{N_{d}} \mathrm{e}^{i(\vartheta_j-\gamma)},
\end{equation}
where $N_d$ is the number of oscillators in the dust.
The modulus $ R(t) \in \left [ 0 , 1 \right ] $  quantifies the degree of synchrony:
in the continuum limit $R \approx 0$ means that the dust is distributed in an asynchronous state, while $R=1$ 
implies a full phase synchronization. 
Under the assumption of a single cluster, Eq.~\eqref{dynamics-eq} can be expressed as
\begin{equation}
m\ddot{\vartheta_i} + \dot{\vartheta_i} =  Kf_{cl}\sin(\Psi-\vartheta_i-\gamma) +
  Kf_{d}R \sin(\Phi-\vartheta_i)\,,\label{globaldynamics-eq}
\end{equation}
where $\Psi$ is the phase of the cluster, while $f_{cl}\equiv 1-f_d$ represents
the fraction of oscillators therein.
Eq.~(\ref{globaldynamics-eq}) describes the evolution of a modulated oscillator, the modulation being 
determined by the 
three time-dependent fields $R$, $\Phi$ and $\Psi$ ~\cite{psi}.
Depending on the initial condition, the oscillator may either: (i) collapse
onto the cluster; (ii) converge towards the dust of unsynchronized oscillators.
First-hand information on the stability of these two regimes can be extracted from the
linearized equations,
\begin{equation} 
m\delta \ddot{\vartheta} + \delta\dot{\vartheta} = - K \left [f_{cl}\cos(\Psi-\vartheta-\gamma) +
  f_{d}R \cos(\Phi-\vartheta)\right ]\delta \vartheta \,,\label{eq:transverse}
\end{equation}
where we have dropped the no-longer necessary subindex $i$.
Any stationary regime is characterized by two Lyapunov exponents: their sum is equal to $-1/m$, which 
quantifies the overall degree of dissipation. A stable cluster is identified by the presence of two negative
Lyapunov exponents: this is the extension of a fixed point to the case of a time-dependent (non-autonomous) dynamical
system. The dust, instead, is identified by one positive exponent: this is an instance of a time-dependent
chaotic regime. In typical regimes, the positive exponent is approximately equal to 0.02. 

Simulations show that a light dust (small $f_d$) does not self-sustain. In this regime, 
isolated clusters appear that are linearly unstable:
they ``emit" oscillators which are eventually absorbed by the dust itself.

For intermediate $f_d$ values ($f_{cl}\gtrsim 0.5$), the 
chimera state is stationary: oscillators do neither migrate towards the dust, nor are they absorbed by the clusters.
In this bistable regime, a probe oscillator, guided by the mean field
without influencing it, may, depending on the initial condition,
collapse onto either the dust or the cluster.
The two basins of attraction are separated by the stable manifold of a second unstable cluster.
While it is computationally hard to reconstruct directly this manifold
which fluctuates, we can offer a glimpse of its structure, by proceeding as follows.
An ensemble of probe oscillators is initialized inside a box encircling the
cluster and let evolve for a long but finite time $t_e$.
The final Euclidean distance between each probe oscillator and the cluster reveals
a fractal intertwining of the two basins of attraction (see Fig.~\ref{fig:kuramoto}(b)).

\begin{figure}
\includegraphics[width=0.5\textwidth,clip]{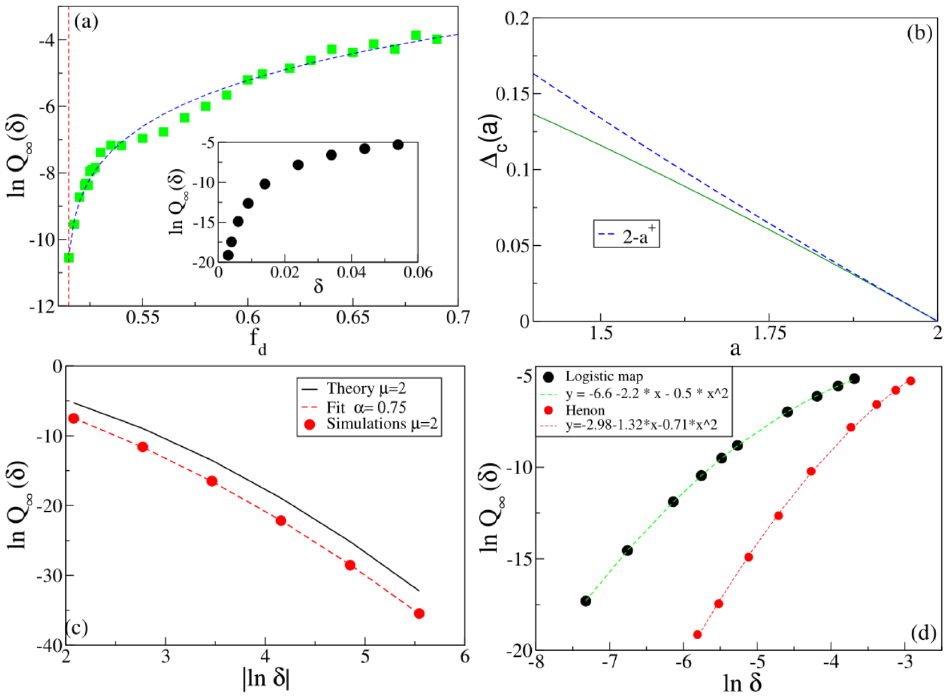}
\caption{(a) Kuramoto model: escape probability towards the cluster as a function of the initial value of $f_d$. 
Each value is obtained by averaging over 10 different realisations each of length $t=2000000$. 
The dashed line is a fit with Eq.~(\ref{eq:theoryf}).
In the inset, the escape probability vs the distance from the critical point is reported for the H\'{e}non map. 
(b) Critical noise amplitude vs the average value of the parameter $a$ for the logistic map. 
(c) Escape probability vs the logarithm of the distance from criticality for the linear stochastic model of the GZ dynamics.
(d) Scaling behavior of the escape probability for the logistic and H\'{e}non map. 
Black (red) dots represent simulation data for the logistic (H\'{e}non) map, while the green (red) dashed line
represent the corresponding scaling behavior estimated by following Eq. (7).}
\label{fig:crisis_log}
\end{figure}

By further increasing $f_d$, an intriguing instability sets in:
oscillators sporadically leave the dust, eventually landing on a single cluster.
One such episode is represented in the inset of Fig.~\ref{fig:kuramoto}(a), where we plot the 
evolution of the instantantaneous Euclidean distance of the center of mass of the dust (red curve) 
together with the maximal (blue curve) and minimal (black curve) distance,
from the cluster. There, we see a sudden approach of the dust to the cluster 
accompanied and followed by the loss of one (or more) oscillators.
More quantitatively, there exists a critical fraction $\tilde f_d \approx 0.515$ above which
the dust becomes metastable. Upon converging to $\tilde f_d$ from above, the escape rate 
from the dust progressively vanishes (see Fig.~\ref{fig:crisis_log}(a)).
Altogether, this is the scenario of a crisis in a regime where the 
dust (the attractor undergoing the transition) is time dependent as well as its basin of attraction. 

\paragraph{Simpler models.}
Now, we consider two stochastically-modulated systems, 
affected by a finite noise to simulate a deterministic chaotic forcing. 
The first model is the H\'enon map,  $y_{n+1} = a_n - y_n^2 + bx_n$, where the control parameter 
$a_n$ is a uniformly distributed ($a_n \in [a-\Delta,a+\Delta]$) $\delta$-correlated noise.
For $a = 1.4$, $b=0.3$, and $\Delta=0$ the map generates the standard H\'enon attractor.
If $a$ is increased above $a_c= 1.42692111 \ldots$, 
the invariant measure crosses the stable manifold of the fixed point
$y^* = (b-1 - \sqrt{(1-b)^2+4a)}/2$, thereby escaping from the basin of attraction. This is a standard crisis.
If we set $\Delta = 0.03$ and progressively increase $a$, the
first escapes from the attractor occur for $a=1.366$, i.e. when the maximum value is $a_{max} = 1.396$, 
below $a_c$. This means that the transition is not simply determined by 
the fluctuations above the critical value of the noiseless system.
The dependence of the outgoing flux from the attractor on the parameter $a$ can 
be appreciated in the inset of Fig.~\ref{fig:crisis_log}(a), where we see that the
scenario is qualitatively very similar to that of the Kuramoto model (see the body of the same figure).

Next we analyse a yet simpler system: the logistic map, $x_{n+1} = a_n - x_n^2$.
In the absence of fluctuations, the basin of attraction is the interval $[x^-,x^+]$, where
$x^- \equiv (-1 - \sqrt{1+4a})/2$ is the negative fixed point, while $x^+ = a$ is the maximum of the map.
Above $a_c=2$, $x^+$ is mapped to the left of $x^-$ so that the trajectory escapes to infinity.

If we let $a_n$ fluctuate in the interval $[a-\Delta_c,a+\Delta_c]$, the escape can happen if the
minimum possible value of $x_n$ at time $n$ (the iterate of the maximum $a_{n-1}$ at the previous
time) is smaller than the leftmost position of the fixed point $x^-_{n}$ at time $n$.
Mathematically,
\[
a-\Delta_c - (a+\Delta_c)^2 = \left( -1-\sqrt{1+4(a-\Delta_c)}\right)/2
\]
where $\Delta_c(a)$ represents the minimal amplitude of the noise such that escapes from the attractor can occur.
The curve is graphically plotted in Fig.~\ref{fig:crisis_log}(b) (green curve): there we see that
in the limit $a=2$, $\Delta_c=0$, we recover the well known behavior of the deterministic logistic map.
Interestingly, we also see that for $a<2$ the maximum value of $a$ ($a^+=a+\Delta_c$) is always strictly
smaller than 2 (see the Supplemental Material \cite{supplemental} for the complete derivation of the formula for $a^+$), showing that, analogously to the
H\'enon map, the presence of fluctuations lowers the critical point. This is evident when looking at the dashed blue curve, where $2-a^+$ is plotted,
in Fig.~\ref{fig:crisis_log}(b).

The average escape probability for $a=1.95$, is reported in Fig.~\ref{fig:crisis_log}(d)
versus $\delta = \Delta -\Delta_c$ ($\Delta_c = 0.01242$).
The observed scaling behavior is explained in the next paragraph.

\paragraph{Scaling behavior.}
We start decomposing the dynamics around criticality into two regimes: 
(i) a standard chaotic phase (CP) in the bulk of the attractor;
(ii) a grey zone (GZ) between the minimal and the maximal position of the ``fixed" point $x^-$ which may end up with
either a final expulsion or a re-injection into the CP. This regime can also be seen as a stochastic motion in
the vicinity of a random saddle.
A linear stochastic model of the GZ dynamics suffices to determine the scaling behavior of the escape times.
We assume that the phase point $y_n$ obeys the following map,
\begin{equation}
y_{n+1} = (y_n-\sigma_n)\mu + \sigma_n \; .
\label{eq:model}
\end{equation}
The iteration amounts to an expansion by a factor $\mu$ of the current distance of $y_n$ from a randomly
selected unstable ``fixed" point $\sigma_n \in [0,1]$ (this mimicks the fluctuations of $a_n$).
The GZ is the unit interval $[0,1]$, and
the initial condition $y_1$ is uniformly distributed within $[1-\delta,1]$, 
where $\delta$ represents the distance from criticality.
Finally, a trajectory terminates when either $y_n<0$, meaning that the point is
unavoidably expelled from the former attractor, or $y_n>1$, meaning that
it is reinjected in the bulk.

The probability density $P_n(y)$ to lie in $[y,y+dy]$ at time $n$
satisfies the Frobenius-Perron equation
\begin{equation}
 P_{n+1}(y) =  \frac{1}{\mu} \int_0^1 d\sigma P_n[y/\mu +\sigma(1-1/\mu)] \; .
 \label{eq:Frobeniusoperator}
\end{equation}
Hence, the probability to escape from the attractor within the first $n$ iterates is
\[
 Q_n(\delta)= \sum_{k=1}^{n} \int_{1-\mu}^0 P_k(y) dy
\]
where the lower limit of the integral is the minimum attainable $y$-value (as from Eq.~\eqref{eq:model}).
We are interested in $Q_\infty(\delta)$.
It is obvious that $Q_\infty(\delta)<Q_n(\delta)+G_n(\delta)$, where 
\begin{equation}
 G_n(\delta) = \int_{0}^1 P_n(y) dy 
\end{equation}
denotes the probability to be in the GZ after $n$ iterates.

A trajectory starting close to 1 cannot initially escape on the left, no matter the values taken by $\sigma_n$.
It can do so, only after $M$ iterates when $\mu^M\delta \ge 1$.
Equivalently, $Q_n=0$ for $n\le M = -\ln \delta /\ln \mu$ and we can conclude that
$Q_\infty(\delta) < G_M(\delta)$.

Interestingly, $P_n$ (and $G_n$) can be analytically estimated for $n\le M$.
It can indeed be verified (see \cite{supplemental}) that
\begin{equation}
\label{eq:piter}
P_n(y) = K_n(y-1+\mu^n\delta)^n  \qquad y \ge 1- \mu^n\delta
\end{equation}
and is zero otherwise, where
\[
K_{n+1} = \frac{K_n}{\mu-1} \frac{1}{(n+1)\mu^{n+1}}.
\]
By solving this recursive relation for the initial condition $K_0 = 1/\delta$, we obtain
\[
K_{n} =  \left[ \delta(\mu-1)^nn!\mu^{n(n+1)/2} \right ]^{-1}.
\]
In virtue of Eq.~(\ref{eq:piter}), the probability $G_n(\delta)$ is therefore (for $n\le M$)
\[
G_n(\delta) = K_n \int_{1-\mu^n\delta}^1 dy (y-1+\mu^n\delta)^n = \frac{K_n(\mu^n\delta)^{n+1}}{n+1}
\]
so that
\begin{equation}
G_n(\delta) = \left(\frac{\delta}{\mu-1}\right)^n \frac{\mu^{n(n+1)/2}}{(n+1)!}.
\end{equation}
By invoking the Stirling approximation
\[
G_n  \approx \exp \left [n \ln \delta - n \ln \mu + \frac{n(n+1)}{2} \ln \mu -n \ln n + n \right ] 
\]
and setting $n=M= -\ln \delta/\ln \mu$, we find that up to the first two leading terms in $\delta$
\begin{equation}
G_M \approx \exp \left [ -\alpha (\ln \delta)^2 +\beta \ln \delta \right] 
\label{eq:theoryf}
\end{equation}
where $\alpha = 1/(2\ln \mu)$. $G_M(\delta)$ is an upper bound of $Q_\infty(\delta)$.
Its decrease with $\delta$ is slower than exponential but faster than any power law.

In Fig.~\ref{fig:crisis_log}(c), we can compare the theoretical prediction (\ref{eq:theoryf})  (black curve)
with the direct values of $Q_\infty(\delta)$ (red dots), for $\mu=2$.
Obviously, $Q_\infty<G_M$. Since the gap between the two quantities does not increase upon decreasing $\delta$,
we can conjecture that $G_M$, i.e. the probability to be still in the GZ when it becomes possible at all
to escape on the left, represents the leading contribution to $Q_\infty(\delta)$ (for $\delta \to \infty$).
In fact, by using $\alpha$ and $\beta$ as fitting parameters, Eq.~(\ref{eq:theoryf}) provides
a very good reproduction of the numerical data: see the dashed line in Fig.~\ref{fig:crisis_log}(c), obtained
for $\alpha \approx 0.75$, to be compared with the theoretical expectation for $G_M$: $1/(2\ln 2) = 0.721\ldots$.

\begin{figure}
\includegraphics[width=0.5\textwidth,clip]{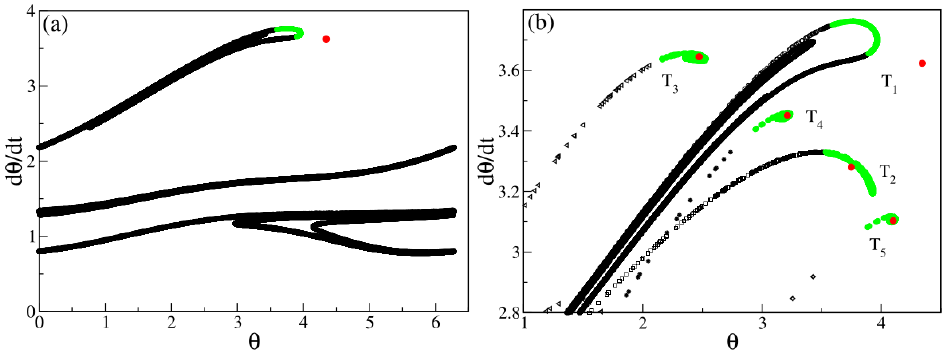}
\caption{(a) Snapshot of the Kuramoto model in the phase plane $(\dot{\vartheta},\vartheta)$. 
The red dot corresponds to the cluster position, while the black and green circles correspond to the dust.
(initially,  $f_{cl}= 0.393$). (b) Enlargement of panel (a) around the cluster position corresponding to
the evolution at time $T_1$. Here are reported enlargements of the same dynamical evolution for successive times $T_2,..,T_5$
(identifiable by different symbols) to characterize the tunnel zone in the cluster growth problem.
}
\label{fig:crisis_GZ}
\end{figure}

\paragraph{Back to dynamical models.} 
Now, we go back to the logistic map.
In Fig.~\ref{fig:crisis_log}(d) we report the data so as to emphasize the quadratic dependence on $\ln \delta$
(see full black dots). A fit in terms of Eq.~(\ref{eq:theoryf}) (with $\alpha$ and $\beta$ as free parameters)
reproduces very well the numerical observations, although now $\alpha \approx 0.5$ differs more
from the theoretical expectation ($\approx 0.4$), the reason being that the quadratic maximum of the logistic map 
induces a singularity in the distribution of initial conditions in the GZ, which is not taken into account in the theory.
In Fig.~\ref{fig:crisis_log}(d) we report also the data for the H\'enon map (full red dots): 
the quality of the fit is again very good, in spite of the two-dimensional character of the phase space.
Hence, the GZ is not an interval containing fluctuating saddle; it 
is a thin corridor covering its stable manifold. Nevertheless, in the small $\delta$ limit,
the scenario is similar, since the relevant trajectories naturally flow towards the saddle point
(observational evidence is offered in~\cite{supplemental}).

Finally, we go back to the Kuramoto model. Here, the collapse onto the cluster is equivalent to the
divergence observed in the logistic map. The single-oscillator dynamics 
is two-dimensional as in the H\'enon map, but now the external modulation plays a double role: it induces a chaotic dynamics
(otherwise impossible in a two-dimensional continuous-time, autonomous dynamical system)
testified by the fractal basin boundary (as visible in Fig.~\ref{fig:kuramoto}(b)) and is responsible for 
the stochastic-like fluctuations of the basin boundary: it is, in fact, well known that mean-field type
models may be characterized by a high- (actually infinite-) dimensional dynamics~(see Ref.~\cite{politi17} and references therein).
Evidence of the pseudo-random oscillations of the order parameter in 
the second-order Kuramoto is given in~\cite{supplemental}.

The direct reconstruction of the GZ is computationally hard, but we can illustrate an escape event.
In Fig.~\ref{fig:crisis_GZ}, we show four different instances of the distribution
of points in correspondence of the escape of some oscillators from the cloud ($10^6$ probe
oscillators have been added to make the scenario clear).
A ``tongue" is initially emitted out by the dust (as a consequence of some
fluctuation); the tip of the tongue reaches the cluster, while the rest pulls back.
The points in the tongue can be interpreted as belonging to the GZ; some of them
(green) eventually leave the attractor (the dust), while others (black) ones are pushed back to the
dust.  
Quantitatively, the numerical values of the escape rate have been fitted with
the theoretical expression Eq.~(\ref{eq:theoryf}). The outcome, reported in
 Fig.~\ref{fig:crisis_log}(a) (see the dashed line, which corresponds to $\alpha = 0.13$), 
reveals again excellent agreement with the raw data.

\paragraph{Conclusions.}
We have shown that the crisis, a typical transition occurring in chaotic attractors, may  also arise in
time-dependent models, although the scaling behavior is very different and the mechanism itself is more
complicate since it is controlled by a tunnelling mechanism induced by the fluctuations of the underlying basin of
attraction.
How to determine $\alpha$? In the simple stochastic one-dimensional model, it is linked to the instability of
the fixed point whose stable manifold determines the boundary of the basin of attraction. More in general,
we can imagine that the fractal dimension of the attractor to enter as well and the correlations probably
play a crucial role. This is left to future work.

\begin{acknowledgments}
AP wishes to acknowledge Celso Grebogi for providing useful information. SO thanks Matthias Wolfrum for useful discussions in the
initial stage of this project. SO received financial support from the National Centre for HPC, Big Data and
Quantum Computing – HPC (Centro Nazionale 01 – CN0000013) CUP B93C22000620006 with particular reference to Spoke 8: In Silico Medicine $\&$
Omics Data.
\end{acknowledgments}


\begin{thebibliography}{68}
\bibitem{Crauel2015}
H. Crauel and P. E. Kloeden, Jahresber. Dtsch. Math. Ver. 117, 173-206 (2015).

\bibitem{Ghil2018}
S. Pierini, M. D. Chekroun, and M. Ghil, Nonlinear Processes in Geophysics 25, 671–692 (2018).

\bibitem{Ramirez2018}
J. Pena Ramirez, A. Arellano-Delgado, and H. Nijmeijer, Phys. Rev. E 98, 012208 (2018).

\bibitem{Clemson2014}
P. T. Clemson and A. Stefanovska, Phys. Rep. 542, 297 (2014).

\bibitem{Grebogi1982}
C. Grebogi, E. Ott, and J. A. Yorke, Phys. Rev. Lett. 48, 1507 (1982).

\bibitem{Checkroun2018}
M. D. Chekroun, M. Ghil, and J. D. Neelin, in {\em Advances in nonlinear geosciences}, edited by A. A. Tsonis (Springer International Publishing, 2018) pp. 1,33.

\bibitem{Sommerer1991}
J. C. Sommerer, E. Ott, and C. Grebogi, Phys. Rev. A 43, 1754 (1991).

\bibitem{Sommerer1991b}
J. C. Sommerer, W. L. Ditto, C. Grebogi, E. Ott, and M. L. Spano, Phys. Rev. Lett. 66, 1947 (1991).

\bibitem{Franaszek1991}
M. Franaszek, Phys. Rev. A, 44(6), 4065.

\bibitem{ReimannI}
P. Reimann, J. Stat. Phys., 82, 1467-1501 (1996).

\bibitem{ReimannII}
P. Reimann, J. Stat. Phys., 85, 403-425 (1996).

\bibitem{Wackerbauer1999}
R. Wackerbauer, Phys. Rev. E, 59(3), 2872 (1999).

\bibitem{kuramoto2002coexistence}
Y. Kuramoto and D. Battogtokh, arXiv preprint condmat/0210694 (2002).
 
\bibitem{abrams2004chimera}
D. M. Abrams and S. H. Strogatz, Phys. Rev. Lett. 93, 174102 (2004).

\bibitem{omel2008chimera}
O. E. Omelchenko, Y. L. Maistrenko, and P. A. Tass, Phys. Rev. Lett. 100, 044105 (2008).

\bibitem{pikovsky2008partially}
A. Pikovsky and M. Rosenblum, Phys. Rev. Lett. 101, 264103 (2008).

\bibitem{laing2009chimera}
C. R. Laing, Chaos: An Interdisciplinary Journal of Nonlinear Science 19 (2009).

\bibitem{olmi2011collective}
S. Olmi, A. Politi, and A. Torcini, EPL 92, 60007 (2011).

\bibitem{scholl2016synchronization}
E. Sch\"{o}ll, The European Physical Journal Special Topics 225, 891 (2016).

\bibitem{zakharova2020chimera}
A. Zakharova, {\em Chimera patterns in networks} (Springer, 2020).

\bibitem{jaros2015chimera}
P. Jaros, Y. Maistrenko, and T. Kapitaniak, Phys. Rev. E 91, 022907 (2015).

\bibitem{olmi2015intermittent}
S. Olmi, E. A. Martens, S. Thutupalli, and A. Torcini, Phys. Rev. E 92, 030901(R) (2015).

\bibitem{olmi2015chimera}
S. Olmi, Chaos: An Interdisciplinary Journal of Nonlinear Science 25 (2015).

\bibitem{psi}
$\Psi$ satisfies Eq.~(\ref{globaldynamics-eq}), in which case, the argument of the first sinusoidal term reduces to $-\gamma$.

\bibitem{supplemental}
See Supplemental Materials for details on the simpler models and high-dimensional chaotic dynamics in Kuramoto oscillators, which includes
Refs. \cite{benettin1980,pikovsky2016}.

\bibitem{politi17}
A. Politi, A. Pikovsky, and E. Ullner, Eur. Phys. J. Special Topics 226, 1791 (2017).

\bibitem{benettin1980}
G. Benettin, L. Galgani, A. Giorgilli, and J.-M. Strelcyn, {\em Meccanica} 15, 9 (1980).

\bibitem{pikovsky2016}
A. Pikovsky and A. Politi, {\em Lyapunov exponents: a tool to explore complex dynamics}, Cambridge University Press (2016).
 
\end{thebibliography}
\end{document}